\documentclass[12pt]{article}
\usepackage{epsf}
\usepackage{amsmath,amssymb}
\usepackage{graphicx}
\usepackage{color}
\usepackage{here}
\usepackage{cite}
\usepackage[colorlinks,citecolor=blue]{hyperref}
\renewcommand{\thefootnote}{\#\arabic{footnote}}

\renewcommand{\thefootnote}{\fnsymbol{footnote}}
\def\thefootnote{\fnsymbol{footnote}}

\makeatletter

\@addtoreset{equation}{section}
\makeatother

\begin{document}

\begin{titlepage}

\begin{center}

\vskip .75in

\bigskip

{\Large \bf Big bang nucleosynthesis and early dark energy  \vspace{2mm} \\ in light of  the EMPRESS $Y_p$ results and the $H_0$  tension}  

\bigskip

\vskip .75in

{\large
Tomo~Takahashi$\,^1$ and Sora~Yamashita$\,^2$ 
}

\vskip 0.25in

{\em
$^{1}$Department of Physics, Saga University, Saga 840-8502, Japan  \vspace{2mm} \\
$^{2}$Graduate School of Science and Engineering, Saga University, Saga 840-8502, Japan
}

\end{center}
\vskip .5in

\begin{abstract}

Recent measurement of the primordial $^4$He abundance $Y_p$ from EMPRESS  suggests a cosmological scenario with the effective number of neutrino species deviated from the standard value  and non-zero lepton asymmetry. We argue that the extension of the standard cosmological model would be more demanded when the Hubble tension is taken into account, in which derived baryon density could be somewhat higher than the standard $\Lambda$CDM framework. We also discuss the issue by assuming early dark energy whose energy density can have a sizable fraction at the epoch of big bang nucleosynthesis. We show that the existence of early dark energy can reduce some tension implied by the EMPRESS $Y_p$ results.

\end{abstract}

\end{titlepage}

\renewcommand{\thepage}{\arabic{page}}
\setcounter{page}{1}
\renewcommand{\thefootnote}{\#\arabic{footnote}}
\setcounter{footnote}{0}

\section{Introduction \label{sec:intro}}

The concordance model of cosmology has now been established, which can successfully explain various cosmological observations almost consistently, which is the so-called $\Lambda$CDM model. However, in recent years, several tensions in the framework of $\Lambda$CDM have been under debate, which may suggest a modification/extension of the concordance $\Lambda$CDM model. One of them is the Hubble tension (the $H_0$ tension) where there is a discrepancy in the values of the Hubble constant $H_0$ measured directly in the local Universe and indirectly from such as cosmic microwave background (CMB) in the framework of $\Lambda$CDM.  More specifically, the Cepheid calibrated supernova distance ladder measurement gives $H_0 = 73.04 \pm 1.04~{\rm km/s/Mpc}$ \cite{Riess:2021jrx}, on the other hand, CMB data from Planck in combination with baryon acoustic oscillation (BAO) measurements infers $H_0 = 67.66 \pm 0.42~{\rm km/s/Mpc}$ \cite{Planck:2018vyg}, which are discrepant at  $4.8 \sigma$ level. Even without CMB data, the tension between direct and indirect measurements still persists \cite{Schoneberg:2019wmt,Okamatsu:2021jil,Schoneberg:2022ggi} and indeed other direct and indirect measurements also show a similar tendency for each category (see, e.g., \cite{DiValentino:2021izs,Perivolaropoulos:2021jda} for a review), which motivates many works to resolve the tension and pursue models beyond the standard $\Lambda$CDM (see, e.g., \cite{DiValentino:2021izs,Schoneberg:2021qvd} for a review). 

Another issue has appeared recently from the measurement of the primordial $^4{\rm He}$ abundance $Y_p$ by Extremely Metal-Poor Representatives Explored by the Subaru Survey (EMPRESS) which obtained $Y_p = 0.2379^{+0.0031}_{-0.0030}$ \cite{Matsumoto:2022tlr}. This value is somewhat smaller than the previously obtained ones such as Aver et al.,  $Y_p = 0.2449 \pm {0.0040}$ \cite{Aver:2015iza}, Hsyu et al., $Y_p = 0.2436^{+0.0039}_{-0.0040}$ \cite{Hsyu:2020uqb}, and Kurichin et al., $Y_p = 0.2462 \pm 0.0022$ \cite{Kurichin:2021ppm}.   Actually, by combining the measurement of the deuterium abundance $D_p$ from Cooke et al. \cite{Cooke:2017cwo}, $D_p = (2.527 \pm 0.030) \times 10^{-5}$, the EMPRESS results give constraints on the effective number of neutrino species $N_{\rm eff} = 2.41^{+0.19}_{-0.21}$ and the baryon-to-photon ratio $\eta \times 10^{10} = 5.80^{+0.15}_{-0.13}$ \cite{Matsumoto:2022tlr} in the framework of $\Lambda$CDM+$N_{\rm eff}$ model. The value of $N_{\rm eff}$ deviates from the standard one\footnote{
Although we use $N_{\rm eff} = 3.046$ \cite{Mangano:2005cc} as a reference value for the standard case, recent precise calculations give $N_{\rm eff} = 3.044 - 3.045$ \cite{deSalas:2016ztq,EscuderoAbenza:2020cmq,Akita:2020szl,Froustey:2020mcq,Bennett:2020zkv}.
} and the baryon-to-photon ratio is slightly smaller than that obtained from Planck in the framework of the $\Lambda$CDM model. This may raise another tension in the standard cosmological model, which is referred to as the ``Helium anomaly" in some literature. 

Actually when one introduces a non-zero chemical potential for electron neutrino $\mu_{\nu_e}$ which is commonly characterized by the degeneracy parameter  $\xi_e = \mu_{\nu_e} / T$ with $T$ being the neutrino temperature,  one obtains $\xi_e = 0.05^{+0.03}_{-0.03}$ and $N_{\rm eff} = 3.22^{+0.33}_{-0.32}$  \cite{Matsumoto:2022tlr} with a Gaussian prior for the baryon-to-photon ratio $\eta \times 10^{10}= 6.132 \pm 0.038$ motivated by the Planck result in the $\Lambda$CDM framework \cite{Planck:2018vyg}\footnote{
See \cite{Seto:2021tad} for the implication of non-zero lepton asymmetry and extra radiation for the $H_0$ tension.
}. This suggests a non-zero lepton asymmetry,  and its implications have been discussed in \cite{Kawasaki:2022hvx,Burns:2022hkq,Borah:2022uos,Escudero:2022okz}.  Instead of invoking a non-zero lepton asymmetry, one can also envisage a model with modified gravity \cite{Kohri:2022vst} to resolve the helium anomaly. In any case, the EMPRESS result may indicate that we need a model beyond the standard paradigm and pose yet another tension in cosmology. 

Actually, as we will argue in this paper when the Hubble tension is taken into account, the above EMPRESS result would imply a non-standard scenario even more.  In many attempts to explain the $H_0$ tension, it can be resolved in such a way that  the value of $H_0$ derived indirectly from the data of CMB, BAO and so on gets higher and becomes close to the one from the direct measurements. Indeed  the baryon density obtained in such model frameworks  tends to be  affected and becomes higher than the value in the $\Lambda$CDM case,  which makes the above-mentioned discrepancy more severe. We discuss this issue quantitatively by investigating the fit to the EMPRESS $Y_p$ result in combination with deuterium abundance from \cite{Cooke:2017cwo} with the prior for baryon density suggested by the $H_0$ tension.

Moreover, we also argue that one can reduce the degree of the helium anomaly by extending the standard model by introducing early dark energy (EDE) model \cite{Poulin:2018cxd} which has  been extensively discussed in the context of the $H_0$ tension (for various works of EDE, see, e.g., \cite{DiValentino:2021izs}). Although the EDE model we consider in this paper has an energy scale different from the one introduced to resolve the $H_0$ tension, the behavior is quite the same in the sense that the energy density of EDE gives a sizable contribution to the total one during the epoch of BBN. Indeed, as we discuss in this paper, we do not need to assume non-zero lepton asymmetry and/or non-standard value of $N_{\rm eff}$ by assuming the existence of EDE in some cases to relax the helium anomaly.

The structure of this paper is as follows. In the next section, we discuss the implications of the Hubble tension to BBN,  particularly focusing on its consequences on the EMPRESS $Y_p$ results through the value of baryon density suggested by models which may resolve the $H_0$ tension. Then in Section~\ref{sec:EDE}, we argue that, by introducing EDE  whose energy fraction becomes relatively large during the BBN era,  the anomaly implied by the EMPRESS results can be alleviated. In the final section, we conclude our paper.

\section{Implications of the Hubble tension to BBN \label{sec:H0}}

In this section, we discuss the implication of the $H_0$ tension to BBN, especially focusing on the effect of the prior on baryon density suggested by the tension. In Table~\ref{tab:eta_H0}, we list the constraints on the baryon density $\Omega_bh^2$ in the $\Lambda$CDM and example models proposed to resolve the $H_0$ tension.  As such example models, most of them are taken from among those referred as ``successful"  models as a possible solution to the $H_0$ tension in \cite{Schoneberg:2021qvd} based on the criteria of the Gaussian tension, the difference of the maximum a posteriori and the Akaike information criterium (see Table~1 in \cite{Schoneberg:2021qvd}). We have taken the constraints on $\Omega_bh^2$ from the references cited in the last column of Table~\ref{tab:eta_H0}.  We should note that the dataset used in the analysis of those references differs, and currently, there is no consensus on which model is more plausible, and hence the values quoted in the table can be regarded as just a reference.  Nevertheless, we can clearly see that,  in models proposed as a solution to the $H_0$ tension, the value of $\Omega_bh^2$ tends to be  higher than the one in the $\Lambda$CDM model, which shows that the $H_0$ tension would have yet another impact on cosmology\footnote{
Other implications of the $H_0$ tension to other aspects of cosmology have been discussed such as cosmological bounds on neutrino masses \cite{Sekiguchi:2020igz} and the scale dependence of primordial power spectrum \cite{Ye:2021nej,Takahashi:2021bti,Jiang:2022uyg,Ye:2022efx}. 
}. 

Actually, this is somewhat expected since the correlation between $\Omega_bh^2$ and $H_0$ in the position and height of acoustic oscillations in CMB angular power spectrum has already been known (see, e.g., \cite{Ichikawa:2007js,Ichikawa:2008pz}) although it becomes somewhat involved in models where the $H_0$ tension can be solved. To resolve the tension, one needs to reduce the sound horizon at recombination $r_s$, but also has to keep CMB angular power spectra almost intact to obtain a good fit to CMB data (and also to BAO).  For example, in time-varying electron mass model \cite{Sekiguchi:2020teg}, one can reduce $r_s$ by slightly raising the electron mass $m_e$ at recombination epoch, however, to keep a good fit to CMB,  the baryon density $\Omega_bh^2$ should also be increased in accordance with the change of $m_e$  as $\delta m_e / m_e = \delta (\Omega_bh^2)/\Omega_bh^2$ (see Ref.~\cite{Sekiguchi:2020teg} for detailed discussion).  Also in the early dark energy model, in order to increase the value of $H_0$,  one needs to shift $\Omega_bh^2$ to a higher value to have the CMB angular power spectrum almost the same, which can be realized by keeping the acoustic scale and angular damping scale almost unchanged  (for details, see, e.g., \cite{Ye:2021nej}).

\begin{table}
  \centering
  \begin{tabular}{|l|c|c|c|c|c|}
\hline
Model  & $100 \Omega_b h^2$  & $\eta_{10}$ &  $H_0$ & Dataset  & Ref. \\ \hline 
$\Lambda$CDM   & $2.242 \pm 0.014$  &   $6.14\pm 0.038 $ & $67.66 \pm 0.42$ 
& (a) & \cite{Planck:2018vyg} \\
Varying $m_e$+$\Omega_k$  & $2.365^{+0.033}_{-0.037} $ & $6.48^{+0.090}_{-0.101}$ &  $72.84^{+1.0}_{-1.0}$
&(b)  & \cite{Sekiguchi:2020teg}  \\ 
Early dark energy ($\phi^4$+AdS)  & $2.346^{+0.017}_{-0.016} $ & $6.42^{+0.047}_{-0.044}$ & $72.64^{+0.57}_{-0.64}$
& (b) & \cite{Ye:2020btb}  \\ 
Early dark energy (axion type)  & $2.285 \pm 0.021$  & $6.26 \pm 0.057 $ & $70.75^{+1.05}_{-1.09}$
& (c)   & \cite{Hill:2020osr}  \\ 
New early dark energy  & $2.292^{+0.022}_{-0.024} $ & $6.27^{+0.060}_{-0.066}$ & $71.4^{+1.0}_{-1.0}$
& (d) & \cite{Niedermann:2020dwg}  \\ 
Early modified gravity  & $2.275 \pm 0.018 $ & $6.23 \pm 0.049$ & $71.21 \pm 0.93$
&(e)  & \cite{Braglia:2020auw}  \\ 
Primordial magnetic field  & $2.266 \pm 0.014 $ & $6.20 \pm 0.038$ & $70.57 \pm 0.61$
& (f) & \cite{Jedamzik:2020krr}  \\ 
Majoron  & $2.267 \pm 0.017 $ & $6.21 \pm 0.047$ &  $70.18 \pm 0.61$
& (a) & \cite{Escudero:2021rfi}  \\ 
\hline
\end{tabular} 
  \caption{ \label{tab:eta_H0} 
  Constraints on the baryon density $\Omega_bh^2$ and its corresponding baryon-to-photon ratio $\eta_{10}$ in models proposed to resolve the $H_0$ tension. For the dataset shown in the 5th column (a)--(f) denotes: (a)~Planck~2018 + BAO, (b)~Planck 2018 + BAO + SN + $H_0$,  (c)~Planck 2018 + BAO + SN + $H_0$ + RSD + DES,   (d)~Planck 2018 + BAO + SN + BBN + $H_0$, (e)~Planck 2018 + SN + $H_0$ and (f)~Planck 2018 + BAO + SN + DES + $H_0$.  Here ``Planck 2018" refers to Planck 2018 TT, TE, EE + lensing,  and ``$H_0$" indicates that the analysis adopts the $H_0$ prior with the value motivated by a direct local measurement such as the one from distance ladder observations.  For details of the analysis, see the references shown in the last column. We should note that most analyses quoted in the table include the $H_0$ prior, and hence some caution should be taken when interpreting the value of $H_0$ obtained in the analyses.}
\end{table}

When one discusses BBN, the baryon density is usually quoated by the baryon-to-photon ratio $\eta =   n_b / n_\gamma $ with $n_b$ and $n_\gamma$ being the number densities of baryon and photon. The conversion factor from the baryon density $\Omega_b h^2$ to the baryon-to-photon ratio $\eta_{10} = 10^{10} \, \eta $  is given by \cite{Consiglio:2017pot} (see also, e.g., \cite{Serpico:2004gx,Steigman:2006nf})
\begin{equation}
\label{eq:eta10_ob}
\eta_{10} = \frac{273.279}{1 - 0.007125Y_p}   \left( \frac{2.7255~{\rm K}}{T_{\gamma 0}} \right)^3 \Omega_b h^2 \,,
\end{equation}
where $T_{\gamma 0}$ is the present photon temperature. Precisely speaking, the conversion factor depends on the primordial helium abundance. Although we have adopted the EMPRESS mean value of $Y_p = 0.2379 $ to obtain $\eta_{10}$, the dependence on $Y_p$ is very weak so that its uncertainty can be neglected compared to the error in $\Omega_b h^2$ obtained from CMB, and hence the actual value of $Y_p$ in Eq.~\eqref{eq:eta10_ob}  scarcely affects the following argument.  We also list the value of $\eta_{10}$ derived by using Eq.~\eqref{eq:eta10_ob} for each model in Table~\ref{tab:eta_H0}. As seen from the table, models proposed to resolve the $H_0$ tension tend to suggest a higher value of $\eta_{10}$ compared to the one in $\Lambda$CDM. In some cases, the mean value of $\eta_{10}$ can be high as $\eta_{10} \sim 6.48$ although, in general,  the error is also somewhat larger than that for the $\Lambda$CDM case.

\begin{figure}[t]
\begin{center}
\includegraphics[width=\textwidth]{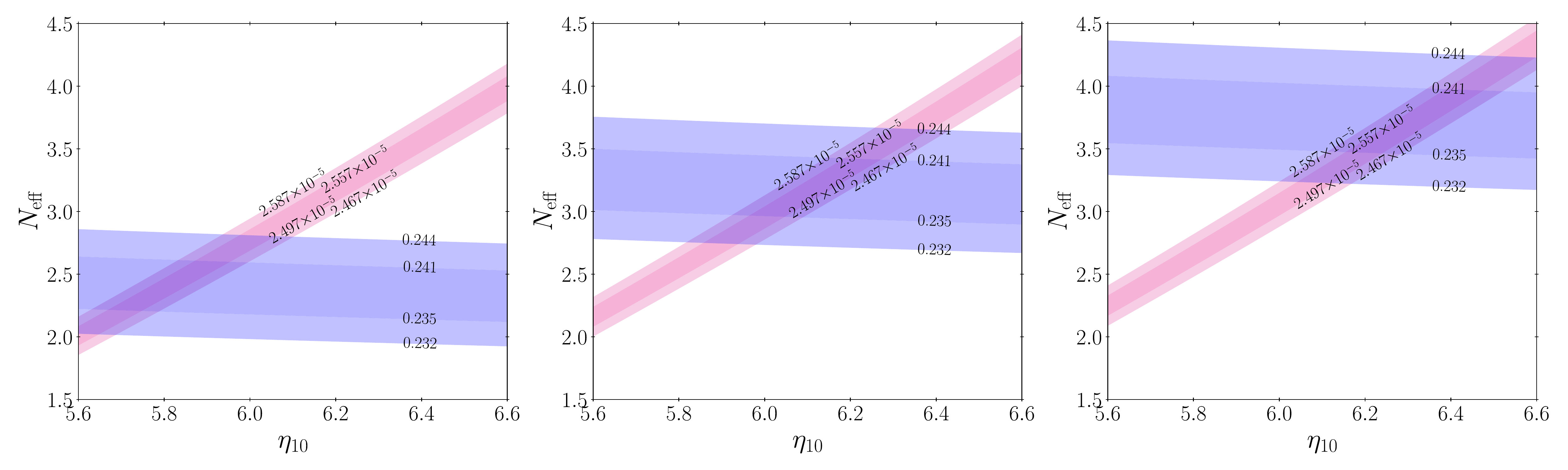}
\end{center}
\caption{\label{fig:eta_xie_Yp_Dp} 
1$\sigma$ and 2$\sigma$ allowed regions for  $Y_p$ \cite{Matsumoto:2022tlr} (light and dark blue regions for 1$\sigma$ and 2$\sigma$) and $D_p$ \cite{Cooke:2017cwo}  (light and dark magenta regions for 1$\sigma$ and 2$\sigma$) in the $\eta_{10}$--$N_{\rm eff}$ plane. We take the neutrino degeneracy parameter as $\xi_e=0$ (left), $0.05$ (middle) and $0.08$ (right). 
}
\end{figure}

Indeed as already argued in \cite{Matsumoto:2022tlr}, the value of $Y_p$ obtained by EMPRESS would imply non-zero lepton asymmetry, which is characterized by non-zero electron neutrino degeneracy parameter $\xi_e$,  and the value of $N_{\rm eff}$ larger than the standard one. This tendency would  become more prominent when the baryon density is high, which can be understood from Fig.~\ref{fig:eta_xie_Yp_Dp} where 1$\sigma$ and 2$\sigma$ allowed region from the measurements of $Y_p$  \cite{Matsumoto:2022tlr} and $D_p$ \cite{Cooke:2017cwo} are shown in the $\eta_{10}$--$N_{\rm eff}$ plane for several values of $\xi_e$.  We use a public code {\tt PArthENoPE} \cite{Pisanti:2007hk,Consiglio:2017pot,Gariazzo:2021iiu} to calculate the helium and deuterium abundances.  As seen from the figure, as $\xi_e$ takes more positive values, $Y_p$ decreases \cite{Kohri:1996ke}, 
which makes higher $\eta_{10}$ more preferred from the combination of the EMPRESS $Y_p$ results \cite{Matsumoto:2022tlr} and $D_p$ from \cite{Cooke:2017cwo}. This indicates that  when a higher value of baryon density is favored from CMB, which can be taken into account in the analysis as a  prior for  $\eta_{10}$ as will be done in the following, a more (positively) non-zero value of $\xi_e$ and  $N_{\rm eff}$ higher than the standard value are more preferred. 

To discuss the implication of a high value of $\eta_{10}$ suggested by the $H_0$ tension in a more quantitative manner, we investigate a constraint in the $N_{\rm eff}$--$\xi_e$ plane from the helium abundance $Y_p$ of EMPRESS \cite{Matsumoto:2022tlr} and deuterium  $D_p$ \cite{Cooke:2017cwo} with the prior for baryon-to-photon ratio~$\eta_{10}$. To obtain the constraint, we calculate $\chi^2$ as 
\begin{equation}
\label{eq:chi2}
 \chi^2 
 = \frac{(Y_p^{\rm obs} - Y_p^{\rm th})^2}{\sigma_{Y_p, {\rm obs}}^2 + \sigma_{Y_p, {\rm sys}}^2 }
 +\frac{(D_p^{\rm obs} - D_p^{\rm th})^2}{\sigma_{D_p, {\rm obs}}^2 + \sigma_{D_p, {\rm sys}}^2 }
 + \frac{(\eta_{10}^{\rm ref} - \eta_{10})^2}{\sigma_{\eta_{10}}^2 } \,,
\end{equation}
where we adopt $Y_p^{\rm obs} =0.2379, \sigma_{Y_p, {\rm obs}} = 0.0031$ \cite{Matsumoto:2022tlr}  and $ D_p^{\rm obs} =  2.527 \times 10^{-5}, \sigma_{D_p}^{\rm obs} = 0.030 \times 10^{-5}$ \cite{Cooke:2017cwo}  for helium and deuterium abundances, respectively. We also include the errors in the theoretical calculation for helium as $ \sigma_{Y_p, {\rm sys}}^2 = (0.0003)^2 + (0.00012)^2$ where the first and second errors  come from the uncertainties on nuclear rate and neutron decay rate with $\tau_n = 879.4 \pm  0.6~{\rm sec}$ \cite{ParticleDataGroup:2020ssz}, respectively \cite{Gariazzo:2021iiu}. For deuterium, we adopt $ \sigma_{D_p, {\rm sys}}^2 = (0.05 \times 10^{-5})^2$ coming from nuclear rate uncertainties \cite{Gariazzo:2021iiu}.  $Y_p^{\rm th}$ and $D_p^{\rm th}$ are theoretically predicted values for given model parameters such as $\eta_{10}, N_{\rm eff}$ and $\xi_e$. When we vary the baryon density $\eta_{10}$ in the analysis, we add the third term to evaluate the value of $\chi^2$.

\begin{figure}[ht]
\begin{center}
\includegraphics[width=9cm]{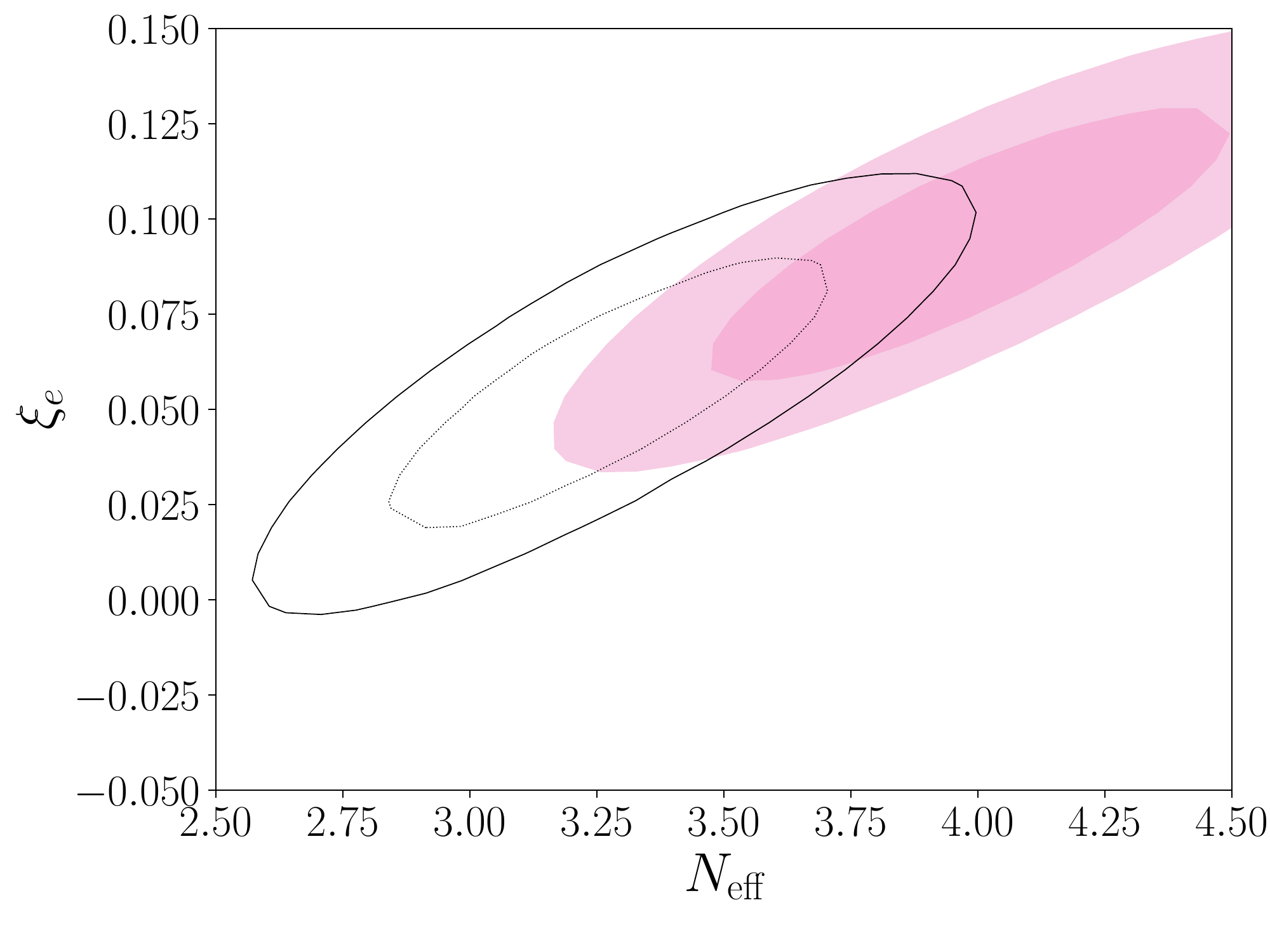}
\end{center}
\caption{\label{fig:Neff_xie_eta} 
1$\sigma$ and 2$\sigma$ allowed region in the $N_{\rm eff}$--$\xi_e$ plane from measurements of $Y_p$ and $D_p$  with $\eta_{10}$ prior. For the $\eta_{10}$ prior, we take $\eta_{10} = \eta_{10}^{\rm ref, 1} \pm \sigma_{\eta_{10,}1}= 6.14 \pm 0.038$ (black dotted and solid lines for 1$\sigma$ and 2$\sigma$) obtained from CMB+BAO \cite{Planck:2018vyg} in the $\Lambda$CDM framework and $\eta_{10} = \eta_{10}^{\rm ref, 2} \pm \sigma_{\eta_{10},2}= 6.4 \pm 0.060$ which is suggested by the $H_0$ tension (light and dark magenta regions for 1$\sigma$ and 2$\sigma$). 
}
\end{figure}

For the baryon density, we consider two priors motivated by the constraints from CMB analysis in the standard $\Lambda$CDM and models proposed to solve the $H_0$ tension and take the following reference values for $\eta_{10}^{\rm ref}$ and $\sigma_{\eta_{10}}$:
\begin{eqnarray}
\label{eq:eta10_ref1}
&& \eta_{10}^{\rm ref,1}  =  6.14, 
\qquad 
\sigma_{\eta_{10},1} = 0.038 \,, \\
\label{eq:eta10_ref2}
&& \eta_{10}^{\rm ref,2}  =  6.40, 
\qquad 
\sigma_{\eta_{10},2} = 0.060 \,.
\end{eqnarray}
For the case 1, which is motivated by the analysis in the standard $\Lambda$CDM framework, we take  $\eta_{10}^{\rm ref, 1}$ and $\sigma_{\eta_{10},1}$ as the same as the $\Lambda$CDM in Table~\ref{tab:eta_H0}.  For the case 2, we take the values motivated from the analysis in models which can resolve the $H_0$ tension. As seen from Table~\ref{tab:eta_H0}, the mean value of $\eta_{10}$  varies from model to model, however one can see the tendency that a higher $\eta_{10}$ is preferred when a model allows a higher value of $H_0$. In order that the value of $H_0$ from indirect measurement such as CMB would be consistent with that obtained from direct measurements, it should be high enough to be close to $H_0 = 73.04 \pm 1.04~{\rm km/s/Mpc}$ \cite{Riess:2021jrx}. Among the values for $H_0$ listed in Table~\ref{tab:eta_H0}, the ones in varying $m_e+\Omega_k$ and early dark energy ($\phi^4$+AdS) models are close to it, in which the mean value of the baryon density is $\eta_{10} = 6.48$ and $6.42$, respectively. Motivated by these values, but here we conservatively take $\eta_{10} = 6.40$ for the reference value of the case 2. The uncertainty of $\eta_{10}$ in those models is somewhat larger than that in the $\Lambda$CDM case. We take $\sigma_{\eta_{10}, 2} = 0.060$ as the reference value, which corresponds to the average value for the uncertainty of $\eta_{10}$ listed in Table~\ref{tab:eta_H0} excluding $\Lambda$CDM (rounded up to the second decimal place).

In Fig.~\ref{fig:Neff_xie_eta}, we show 1$\sigma$ and 2$\sigma$ allowed regions in the $N_{\rm eff}$--$\xi_e$ plane obtained by evaluating $\chi^2$ given in Eq.~\eqref{eq:chi2}. From the figure, one can see that when a higher value of $\eta_{10}$ is assumed for the prior,  more positively non-zero lepton asymmetry (non-zero degeneracy parameter for the electron neutrino $\xi_e$) and larger value of $N_{\rm eff}$ are preferred, which indicates that, in the light of the $H_0$, the EMPRESS result on $Y_p$ gives a more pronounced consequence for $N_{\rm eff}$ and the lepton asymmetry (when the previous measurement of $Y_p$ such as from \cite{Aver:2015iza,Hsyu:2020uqb,Kurichin:2021ppm} is adopted,  the allowed ranges for $N_{\rm eff}$ and $\xi_e$ are closer to the standard ones compared to the case using the EMPRESS $Y_p$).  

Here we discussed the implications of a high baryon density, which may be suggested by the $H_0$ tension, for the EMPRESS $Y_p$ results in the framework where a non-zero lepton asymmetry characterized by $\xi_e$ and a non-standard value of $N_{\rm eff}$ are allowed. Then we showed that a positively large non-zero $\xi_e$ and a larger $N_{\rm eff}$ than the standard case are more preferred. However, we can also consider another framework to discuss the implications of the EMPRESS $Y_p$ results in the light of the $H_0$ tension. As such an example, we will consider early dark energy model in the next section.

\section{Big bang nucleosynthesis with early dark energy \label{sec:EDE}}

Now in this section, we discuss the impact of early dark energy (EDE), which has been intensively studied in the context of the $H_0$ tension, on BBN in the light of the EMPRESS result on $Y_p$.  First we present the EDE model considered in this paper, then we investigate constraints on $\eta_{10}, N_{\rm eff}$ and $\xi_e$ from the EMPRESS $Y_p$  in combination with the measurement of $D_p$ in the existence of EDE.

\subsection{Early dark energy}

EDE models have been  extensively investigated in the context of the $H_0$ tension, whose typical realization is given by a scalar field $\phi$ with a potential, for example, such as $V (\phi)= V_0 \left[ 1 - \cos (\phi /f) \right]^n$, with $V_0$ representing the energy scale,  $f$ being a parameter in the model and $n$ controlling the scaling of its energy density after $\phi$ starts to oscillate.  A general behavior of the energy density of EDE $\rho_{\rm EDE}$ is that, when  $\phi$ slowly rolls on the potential in the early times, $\rho_{\rm EDE}$ is almost constant and acts like a cosmological constant, and then, when the effective mass of $\phi$ becomes the same as the Hubble rate, it starts to oscillate around the minimum of its potential. Around the minimum, the potential can be approximated as $V \propto \phi^{2\alpha}$ and hence, its energy density scales as $\rho_{\rm EDE} \propto a^{-4}$ for $\alpha=2$, $\rho_{\rm EDE} \propto a^{-9/2}$ for $\alpha=3$, and so on. If EDE starts to oscillate around the epoch of recombination and has some energy fraction at the beginning of its oscillation, EDE affects the evolution of perturbations around recombination, and then soon becomes irrelevant to the evolution of the Universe since $\rho_{\rm EDE}$ dilutes away faster than matter, which is a scenario considered in the context of the $H_0$ tension. However, here we discuss a case where EDE can have a sizable energy density fraction at some time during BBN.

In the following, we consider two types of EDE: The first one is essentially the same as that adopted to resolve the $H_0$ tension \cite{Poulin:2018cxd}, in which an EDE component behaves like a cosmological constant at early times, and then its energy density quickly dilutes away at some epoch during BBN.  As mentioned above, the energy scale of EDE considered here is different from that motivated by the $H_0$ tension, such an EDE can be realized by assuming appropriate model parameters even for the same potential as the one adopted to resolve the $H_0$ tension. One could also think of a scenario where two (or more) EDEs are embedded in one framework  like in chain early dark energy model \cite{Freese:2021rjq} in which the Universe experiences multiple first-order phase transitions and some of them act as EDE at BBN and recombination epochs. Another such an example is cascading dark energy \cite{Rezazadeh:2022lsf} where multiple scalar fields can act as dark energy at different eras, which can also accommodates a scenario that  EDEs affect the epochs  of recombination and BBN such that they relax both the $H_0$ tension and the helium anomaly.

Here we just assume an EDE which can have a sizable energy fraction at BBN epoch, and then its energy density dilutes quickly, which we describe its evolution of energy density by adopting the following phenomenological model for simplicity and generality such that the description can capture an essential behavior of the model. We model the evolution of the energy density of the first type of EDE, which we refer to as ``EDE1" in the following, as
\begin{equation}
\rho_{\rm EDE,1} =
\begin{cases}
  \rho_0     &  (T \ge T_t) \,, \\  \\
  \rho_0 \left( \displaystyle\frac{T}{T_t} \right)^n    & (T < T_t)  \,,
\end{cases}
\end{equation} 
where $T$ is the cosmic temperature and $T_t$ is the transition temperature at which the energy density changes its behavior. $n$ is a parameter which describes the scaling of its energy density. Since the Universe is radiation-dominated during BBN, the temperature essentially scales as $T \propto 1+ z \propto 1/a$. $\rho_0$ is assumed to be constant, and hence it represents the vacuum energy before EDE starts to dilute away. We have modified {\tt PArthENoPE} code \cite{Pisanti:2007hk,Consiglio:2017pot,Gariazzo:2021iiu} to include the EDE.  In the calculation, we actually use the energy fraction of EDE at the time of transition, denoted as  $f_{\rm EDE}$,  to control the transition time instead of directly using $T_t$, which is defined by
\begin{equation}
f_{\rm EDE} \equiv \frac{\rho_{\rm EDE} (T_t)}{ \rho_{\rm tot} (T_t) }  = \frac{\rho_{\rm EDE} (T_t)}{ \rho_{\rm EDE} (T_t) + \rho_{rB} (T_t)}  \,,
\end{equation} 
where $\rho_{\rm tot}$ is the total energy density including the EDE component and $\rho_{rB} (T)$ is the sum of energy densities of photon, neutrino, electron and baryon.  In our calculation, we take account of the time variation of $\rho_{rB}$ properly and set $T_t$ for a given $f_{\rm EDE}$. One can approximately evaluate $T_t$ for a given $f_{\rm EDE}$ as $T_t \sim 1~{\rm MeV} \left( (\rho_0/1~{\rm MeV}^4) (1-f_{\rm EDE}) \right)^{1/4}$. Since $\rho_{rB}$ monotonically decreases, but $\rho_{\rm EDE}$ is constant when $T > T_t$, the impact of EDE is  the largest at around $T=T_t$.

The second type of EDE we consider is the one which behaves as a negative cosmological constant until some time during BBN and it quickly settles down to the observed cosmological constant today. Although EDE having a negative energy density seems somewhat contrived or exotic, such a negative cosmological constant has been investigated in the context of the $H_0$ tension \cite{Mortsell:2018mfj,Ye:2020btb,Poulin:2018zxs}, and the tension in BAO observations at $z \simeq 2.4$ between the ones observed from Lyman $\alpha$ forest and the predicted values in $\Lambda$CDM model\footnote{
Although the tension has been suggested as $\sim 2 \sigma  - 2.5  \sigma$ \cite{BOSS:2014hwf,Aubourg:2014yra,duMasdesBourboux:2017mrl}, it is reduced to $1.5 \sigma$ in a recent measurement \cite{duMasdesBourboux:2020pck}. 
} \cite{Wang:2018fng,Dutta:2018vmq}.  Furthermore, some theoretical frameworks motivate a negative dark energy such as bimetric gravity \cite{Fasiello:2013woa,Akrami:2015qga,Konnig:2015lfa}, graduated dark energy \cite{Akarsu:2019hmw,Akarsu:2021fol,Akarsu:2022typ}, everpresent $\Lambda$ \cite{Ahmed:2002mj,Zwane:2017xbg} and so on. In particular, in the everpresent $\Lambda$ model, the energy fraction of negative dark energy can have a sizable contribution to the total one at some time due to its stochastic nature \cite{Ahmed:2002mj,Zwane:2017xbg}. Although one could predict the evolution of such dark energy for a given model, here we assume that the EDE energy density changes  from a negative constant to almost zero (actually to a very small value which can explain the present-day dark energy) at some time during BBN. Since a tiny cosmological constant should be negligible compared to other energy components during BBN epoch,  to study the effect of this second type of EDE, which we refer to as ``EDE2" in the following,  we model the energy density of EDE2 simply as 
\begin{equation}
\rho_{\rm EDE,2} = 
\begin{cases}
  - \rho_0     &  (T \ge T_t) \,, \\  \\
0     & (T < T_t)  \,,
\end{cases}
\end{equation} 
where $\rho_0$ is a constant, whose value represents the energy density of EDE2 at early time.  As in the case of EDE 1, instead of using $T_t$, we in practice use the energy density fraction of EDE $f_{\rm EDE}$ to specify the time when the transition from $\rho_{\rm EDE,2} = -\rho_0$ to  $\rho_{\rm EDE,2} = 0$ occurs in our analysis. Since this kind of EDE can reduce the expansion rate of the Universe at some certain period during BBN,  the study of this type of EDE also gives a general insight into models where the expansion rate diminishes at some particular epoch during BBN.

By assuming two types of EDE described above, we discuss the impact of EDE on the abundance of light elements and its implications for the helium anomaly. In particular, we investigate constraints on the parameters such as $\eta_{10}, N_{\rm eff}$ and $\xi_e$ in the presence of EDE to discuss its impact on BBN, which will be presented in the next section.

\subsection{Impact of EDE to BBN}

\begin{figure}[t]
\begin{center}
\includegraphics[width=16cm]{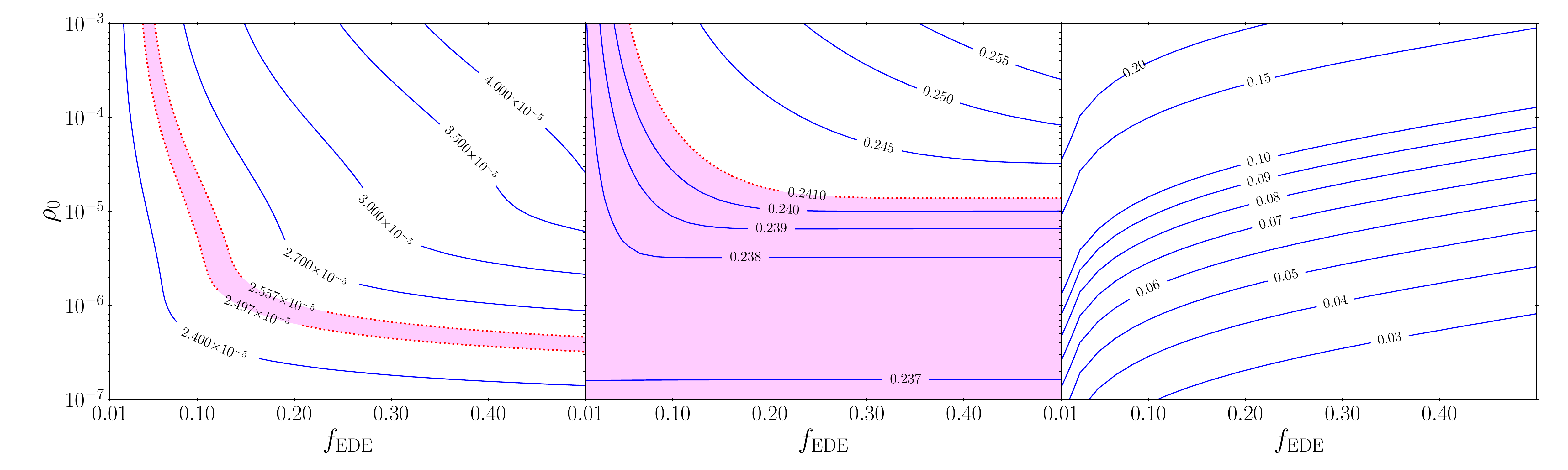}
\end{center}
\caption{\label{fig:EDE1_n4_Yp_Dp_Tt_eta6.14}  
Contours of $D_p$ (left),  $Y_p$ (middle) and $T_t ~[{\rm MeV}]$ (right) in the $f_{\rm EDE}$--$\rho_0$ plane for EDE1 with $n=4$. 1$\sigma$ allowed region for $Y_p$ and $D_p$ is shaded with light magenta.  The value of $\rho_0$ is shown in ${\rm MeV}^4$ unit. In this figure, we take  $\eta_{10} = 6.14, N_{\rm eff} =2.3$ and $\xi_e=0$. }
\end{figure}

\begin{figure}[t]
\begin{center}
\includegraphics[width=16cm]{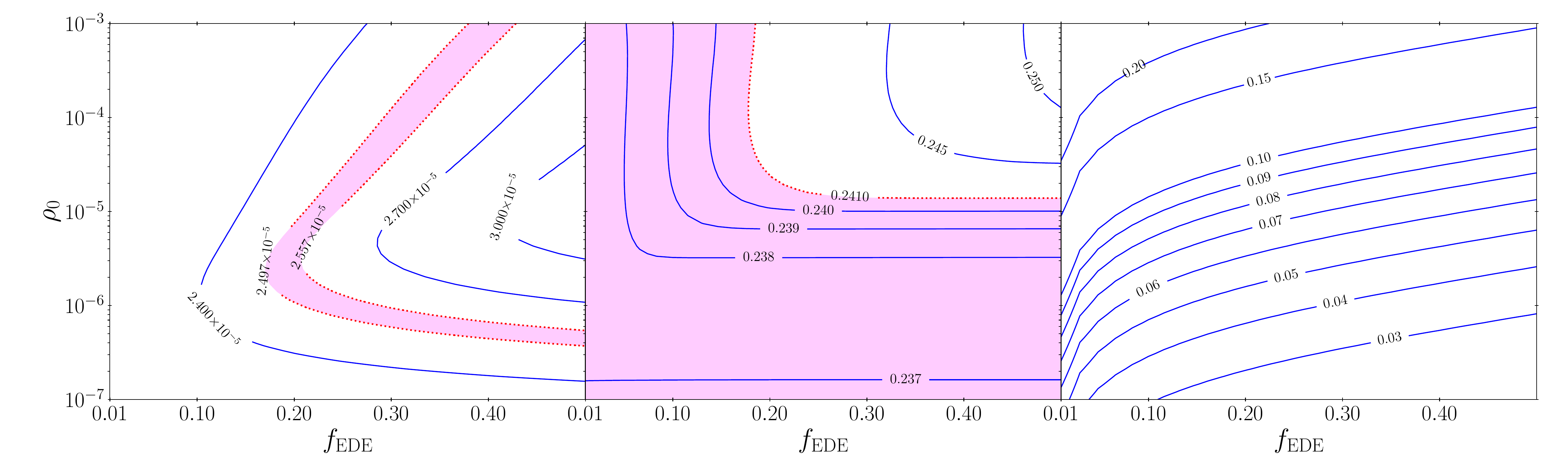}
\end{center}
\caption{\label{fig:EDE1_n6_Yp_Dp_Tt_eta6.14}  
The same as Fig.~\ref{fig:EDE1_n4_Yp_Dp_Tt_eta6.14} except that $n=6$ is assumed in this figure.}
\end{figure}

Here we discuss how the existence of EDE  affects constraints on $\eta_{10}, N_{\rm eff}$ and $\xi_e$ from the abundances of helium and deuterium, especially in the light of the result from EMPRESS on $Y_p$ and the $H_0$ tension.

First we show contours of $D_p, Y_p$ and $T_t$ in the $f_{\rm EDE}$--$\rho_0$ plane for the EDE1 with $n=4$ and $n=6$ cases in Figs.~\ref{fig:EDE1_n4_Yp_Dp_Tt_eta6.14} and \ref{fig:EDE1_n6_Yp_Dp_Tt_eta6.14}, respectively. Other parameters  are taken as  $\eta_{10}=6.14, N_{\rm eff} =2.3$ and $\xi_e=0$. The reason why we take $N_{\rm eff}= 2.3$, which is smaller than the standard value, is that when $N_{\rm eff}=3.046$ and $\xi_e=0$ are assumed, the value of $Y_p$ always larger than the $1\sigma$ upper limit obtained by EMPRESS in the ranges of $f_{\rm EDE}$ and $\rho_0$ shown in the figure, but by lowering the value of $N_{\rm eff}$, $Y_p$ gets smaller so that the 1$\sigma$ allowed range for $Y_p$ becomes visible.   Since the existence of EDE affects the abundance of light elements through the change of the expansion rate of the Universe, in which the freeze-out time of nuclear reactions gets modified, the abundances of helium and deuterium change depending on $f_{\rm EDE}$ and $\rho_0$.  As $f_{\rm EDE}$ or $\rho_0$ increases (i.e., the effects of EDE becomes larger), the expansion rate gets bigger particularly around $T=T_t$, which makes the neutron freeze-out earlier and hence the value of $Y_p$  increases. The same tendency also holds true for deuterium, which can be seen from the left panel of the figure.  It should also be noticed that, in the bottom half  region in the middle panel,  $Y_p$ scarcely changes even when $f_{\rm EDE}$ is increased. Indeed this region corresponds to the one where the transition temperature is lower than $0.07~{\rm MeV}$ as seen from the right panel of the figure.    At around this temperature, the helium abundance almost freezes out, below which the value of $Y_p$ would not be affected even if the expansion rate is changed, i.e., the amount of EDE is irrelevant below $T_t \sim 0.07~{\rm MeV}$. This is the reason why $Y_p$ almost stays constant regardless of the change of $f_{\rm EDE}$.   Compared to $Y_p$,  since the deuterium abundance evolves gradually and does not reach a constant value until late time,  $D_p$ decreases continuously against the changes of $f_{\rm EDE}$ and $\rho_0$ although the response becomes insensitive in the bottom region as in the case of $Y_p$.

Indeed the cases of $n=4$ and $6$ show almost the same tendencies for both $D_p$ and $Y_p$ (contours for $T_t$ are the same for $n=4$ and $6$ since $T_t$ is the temperature when the energy density of EDE1 changes its behavior which is irrelevant to the scaling of $\rho_{\rm EDE1}$ below $T = T_t$).  However, some differences, particularly in $D_p$, appear  in the region above $\rho_0 \gtrsim {\cal O}(10^{-6})~{\rm MeV}^4$. Since the scaling of the energy density of EDE1 in the $n=4$ case is the same as that of radiation, it mimics the effects of $N_{\rm eff}$ below $T=T_t$, on the other hand, when $n=6$, $\rho_{\rm EDE}$ quickly dilutes away and EDE scarcely affect the expansion rate any more. Therefore the difference between the cases with $n=4$ and $6$ only appears when the EDE1 makes a contribution sufficient enough to affect the Hubble expansion rate even for $T <T_t$.  However, allowed overlapping regions between $Y_p$ and $D_p$ are almost the same for the $n=4$ and $6$ cases, and the final constraint also becomes  almost similar. Therefore we only consider the case of $n=6$ for EDE1 in the following argument. It should be noted that the energy density of EDE1 with $n=6$ dilutes faster than radiation and becomes negligible soon below $T=T_t$. Therefore it does not affect the later evolution of the Universe, and  such an EDE would only affect the BBN epoch.

\begin{figure}[H]
\begin{center}
\includegraphics[width=16cm]{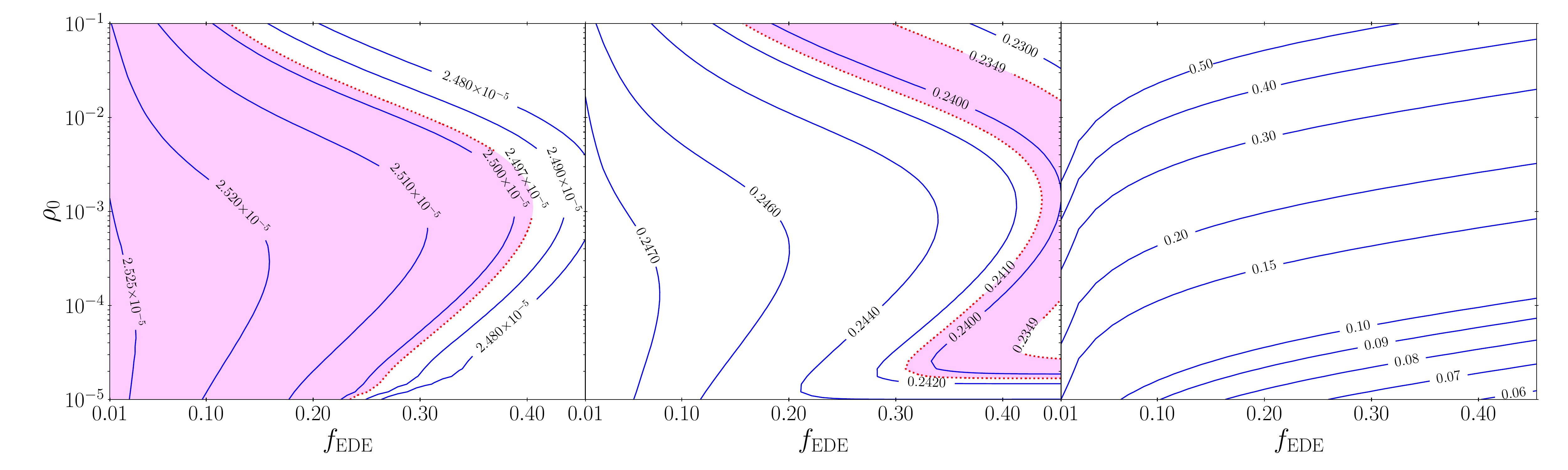}
\end{center}
\caption{\label{fig:EDE2_Yp_Dp_Tt_eta6.14}  
Contours of $D_p$ (left),  $Y_p$ (middle) and $T_t~[{\rm MeV}]$ (right) in the $f_{\rm EDE}$--$\rho_0$ plane for EDE2. 1$\sigma$ allowed region for $Y_p$ and $D_p$ is shown with light magenta. The value of $\rho_0$ is shown in ${\rm MeV}^4$ unit. In this figure, we take  $\eta_{10} = 6.14, N_{\rm eff} =3.046$ and $\xi_e=0$. }
\end{figure}

In Fig.~\ref{fig:EDE2_Yp_Dp_Tt_eta6.14}, we show the case of EDE2. We take $\eta_{10}=6.14$ and $\xi_e=0$ as in Figs.~\ref{fig:EDE1_n4_Yp_Dp_Tt_eta6.14}~and~\ref{fig:EDE1_n6_Yp_Dp_Tt_eta6.14}, but $N_{\rm eff} =3.046$ for this case. Since the EDE2 gives a negative contribution to the total energy density of the Universe, which slows down the Hubble expansion rate, the responses of $Y_p$ and $D_p$ to the changes of $f_{\rm EDE}$ and $\rho_0$ show an opposite tendency to the case of EDE1, i.e., as $f_{\rm EDE}$ and/or $\rho_0$  increase, the values of $Y_p$ and $D_p$ decrease. However, as in the EDE1 case, the value of $Y_p$ is almost unchanged around the bottom right region in the middle panel where the transition temperature is $T_t \lesssim 0.07~{\rm MeV}$. As mentioned above, the helium abundance is almost fixed at this temperature, and hence the change of the expansion rate does not affect the final value of $Y_p$ when $T_t < 0.07~{\rm MeV}$, which is the reason why $Y_p$ stays almost the same as $f_{\rm EDE}$ increases.

\begin{figure}[H]
\begin{center}
\includegraphics[width=16cm]{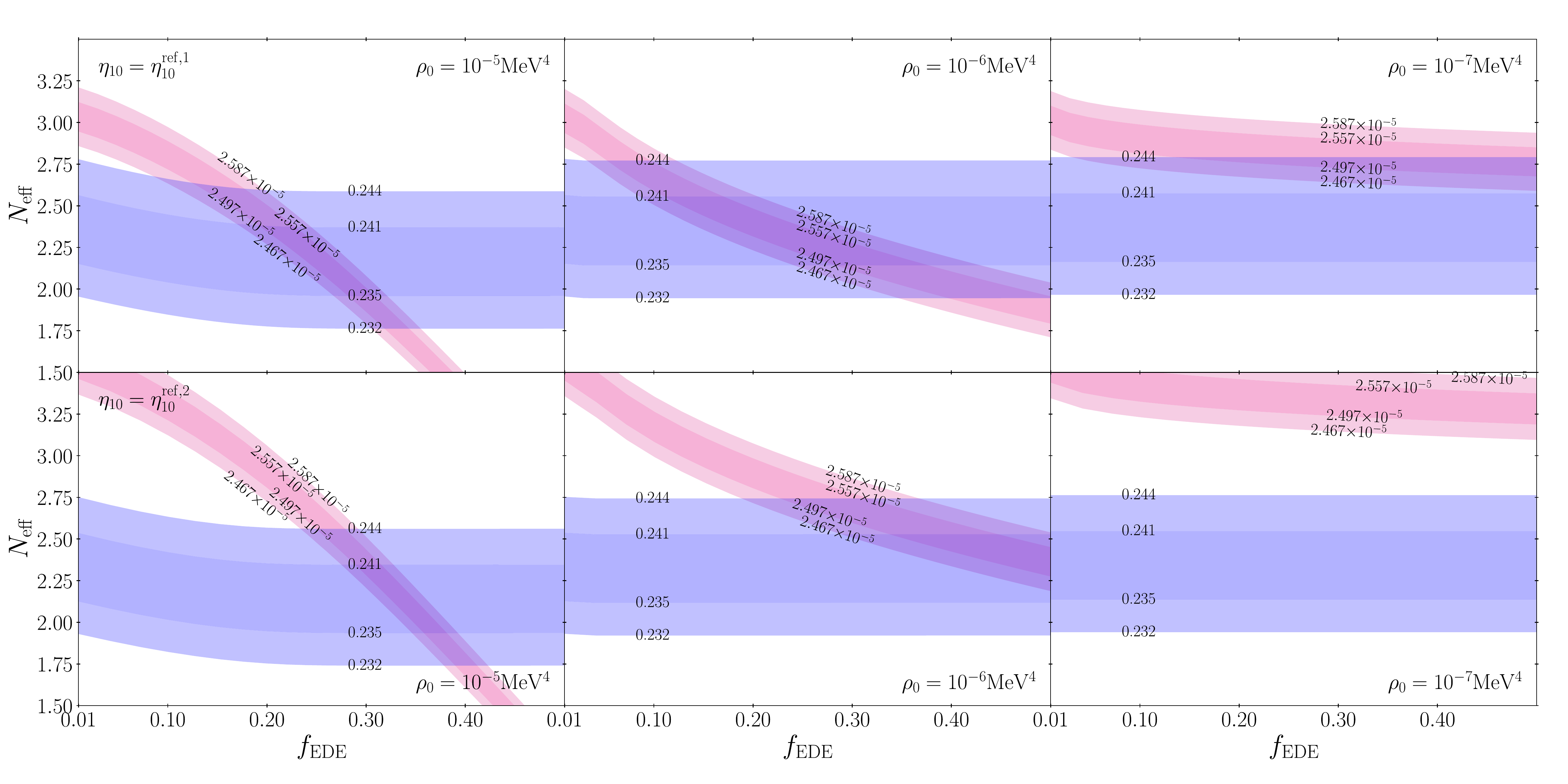}
\end{center}
\caption{\label{fig:YpDp_Neff_fEDE_1} 
1$\sigma$ and 2$\sigma$ allowed ranges from observations of $Y_p$ \cite{Matsumoto:2022tlr} (light and dark blue) and  $D_p$ \cite{Cooke:2017cwo} (light and dark magenta) in the $f_{\rm EDE}$--$N_{\rm eff}$ plane for  EDE1 with $n=6$. The prior for $\eta_{10}$ assumed in the analysis is $\eta_{10} = \eta_{10}^{\rm ref,1}$ (top panels) and $\eta_{10}^{\rm ref,2}$ (bottom panels) as shown in the figure. The value of $\rho_0$ is fixed whose value also appears in the figure. In all panels, the electron neutrino degeneracy parameter is fixed as $\xi_e=0$.}
\end{figure}

\begin{figure}[H]
\begin{center}
\includegraphics[width=16cm]{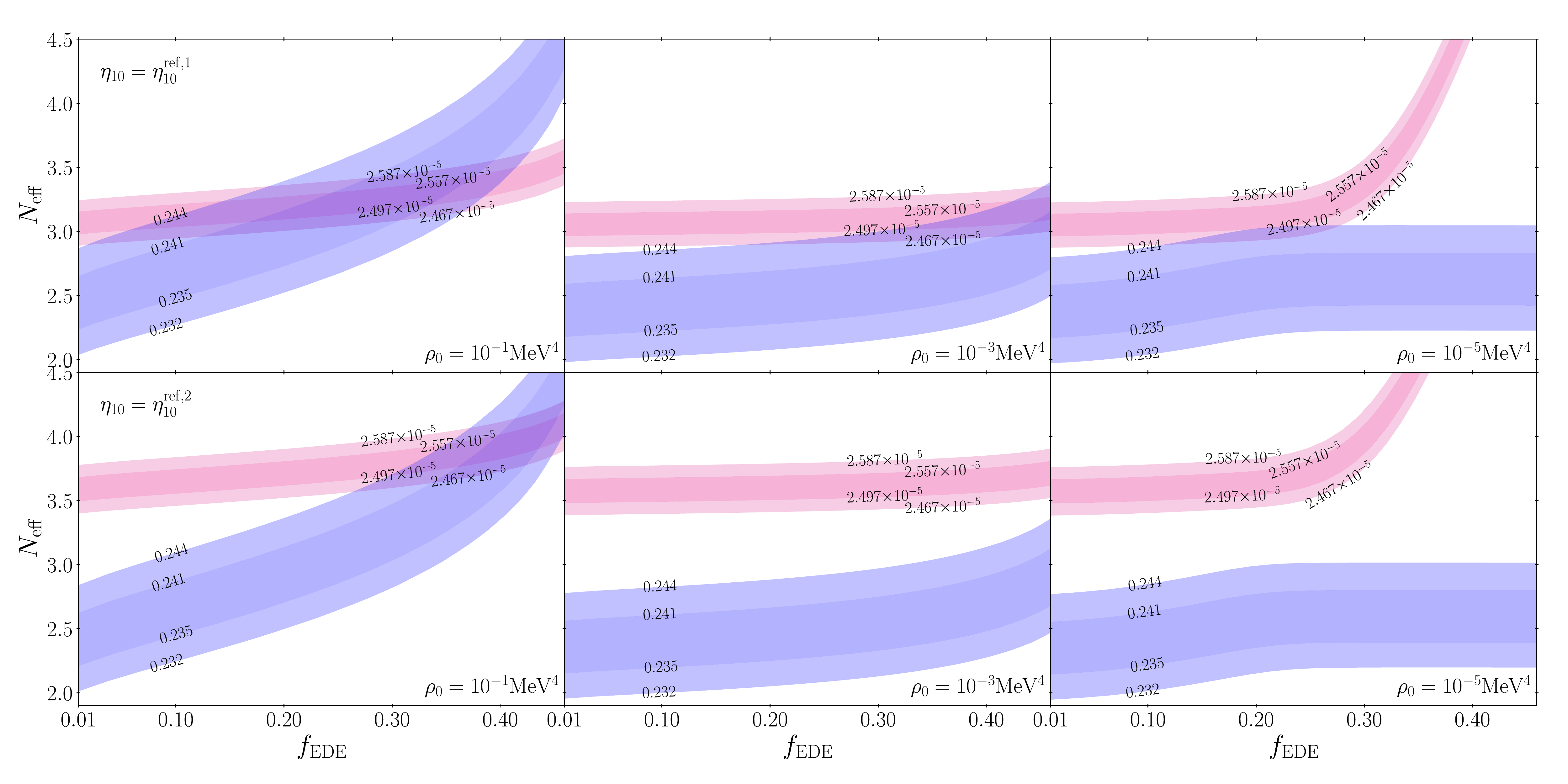}
\end{center}
\caption{\label{fig:YpDp_Neff_fEDE_2} 
1$\sigma$ and 2$\sigma$ allowed ranges from observations of $Y_p$ \cite{Matsumoto:2022tlr} (light and dark blue) and  $D_p$ \cite{Cooke:2017cwo} (light and dark magenta) in the $f_{\rm EDE}$--$N_{\rm eff}$ plane for EDE2. The prior for $\eta_{10}$ assumed in the analysis is $\eta_{10} = \eta_{10}^{\rm ref,1}$ (top panels) and $\eta_{10}^{\rm ref,2}$ (bottom panels) as shown in the figure. The value of $\rho_0$ is fixed whose value also appears in the figure. In all panels, the electron neutrino degeneracy parameter is fixed as $\xi_e=0$.}
\end{figure}

Next we show 1$\sigma$ and 2$\sigma$ allowed ranges from observations of  $Y_p$ \cite{Matsumoto:2022tlr} and  $D_p$ \cite{Cooke:2017cwo} in the $f_{\rm EDE}$--$N_{\rm eff}$ plane in Figs.~\ref{fig:YpDp_Neff_fEDE_1} and \ref{fig:YpDp_Neff_fEDE_2} for the cases of EDE1 and EDE2, respectively.  In the figures, we fix the baryon density to $\eta_{10}^{\rm ref,1}$ and $\eta_{10}^{\rm ref,2}$ as shown in the plots and take several values for $\rho_0$ which are also indicated in the figure. In all cases, we take the electron neutrino degeneracy parameter as $\xi_e=0$. From the figure, one can notice that when there is no EDE (i.e., when $f_{\rm EDE} \rightarrow 0$), there is almost no overlapping region between the allowed ones from the $Y_p$ and $D_p$ measurements in all cases for both EDE1 and EDE2. This is because,  as seen from Fig.~\ref{fig:eta_xie_Yp_Dp},  the allowed regions from $Y_p$ and $D_p$ overlap at $\eta_{10} \sim 5.8$ in the $\eta_{10}$--$N_{\rm eff}$ plane when $\xi_e =0$, but here we fix $\eta_{10}$ to $\eta_{10}^{\rm ref,1}$ and $\eta_{10}^{\rm ref,2}$, which is larger than $\eta_{10} \sim 5.8$. However, as $f_{\rm EDE}$ increases, the overlapping region between the ones allowed by $Y_p$ and $D_p$ measurements appears, which indicates that EDE can improve the fit.

For the case of EDE1 with the value of $\rho_0$ shown in the figure, the transition temperature $T_t$ is smaller than $0.07 ~{\rm MeV}$ in most range of $f_{\rm EDE}$, and hence the value of $Y_p$ scarcely changes even when $f_{\rm EDE}$ is increased as explained above.  It should also be noticed that $N_{\rm eff}$ has to be smaller than the standard value ($N_{\rm eff} = 3.046$) when $\xi_e=0$ to satisfy the EMPRESS $Y_p$ value as seen from the left panel of Fig.~\ref{fig:eta_xie_Yp_Dp}. Therefore the region allowed by the $Y_p$ measurement lies below  $N_{\rm eff} = 3.046$ almost horizontally to the axis of $f_{\rm EDE}$.  Contrary to the behavior of $Y_p$, the deuterium abundance $D_p$ gets smaller as $f_{\rm EDE}$ increases, and hence the overlapping region appears at some value of $f_{\rm EDE}$, which means that we do not need to assume  lepton asymmetry (however, with non-standard value of $N_{\rm eff}$) when the EDE exists. This holds true for both priors of $\eta_{10}^{\rm ref,1}$ and $\eta_{10}^{\rm ref,2}$ although a large fraction of EDE is required when the baryon density is large (i.e., in the case of the prior of $\eta_{10}^{\rm ref,2}$), which can be noticed by comparing the upper and lower panels in Fig.~\ref{fig:YpDp_Neff_fEDE_1}.  When $\rho_0$ is taken to be small,  there is no overlapping region due to the small contribution from EDE, particularly for the prior of $\eta_{10}^{\rm ref,2}$.

In the case of EDE2 shown in Fig.~\ref{fig:YpDp_Neff_fEDE_2},  the responses of $D_p$ and $Y_p$ against the increase of $f_{\rm EDE}$ are opposite to those in the EDE1, which is already presented in Fig.~\ref{fig:EDE2_Yp_Dp_Tt_eta6.14}. However, as in the EDE1 case shown in Fig.~\ref{fig:YpDp_Neff_fEDE_1}, the overlapping region between the allowed ones from the $Y_p$ and $D_p$ measurements appears as $f_{\rm EDE}$ increases.  It should be noticed that the value of $N_{\rm eff}$ at the overlapping region lies around the standard value for the case of the prior of $\eta_{10}= \eta_{10}^{\rm ref,1}$. This means that the existence of EDE2 can help to fit the EMPRESS $Y_p$ result  \cite{Matsumoto:2022tlr}  in combination with the $D_p$ data from  \cite{Cooke:2017cwo} without assuming the deviation of $N_{\rm eff}$ from the standard value and with no lepton asymmetry.  On the other hand, when a higher baryon density prior $\eta_{10} =\eta_{10}^{\rm ref,2}$ is adopted, having an overlapping region becomes a bit difficult, which can be observed from the lower panels in Fig.~\ref{fig:YpDp_Neff_fEDE_2}. Even when such a region exists, the value of $N_{\rm eff}$ is somewhat higher than the standard value.  However, we again remark that no lepton asymmetry is assumed in all cases shown in Fig.~\ref{fig:YpDp_Neff_fEDE_2}. Thus the EMPRESS $Y_p$ result can be well fitted by assuming the existence of EDE without lepton asymmetry although some non-standard value for $N_{\rm eff}$ could be required, particularly when the baryon density is high.

\begin{figure}[t]
\begin{center}
\includegraphics[width=16cm]{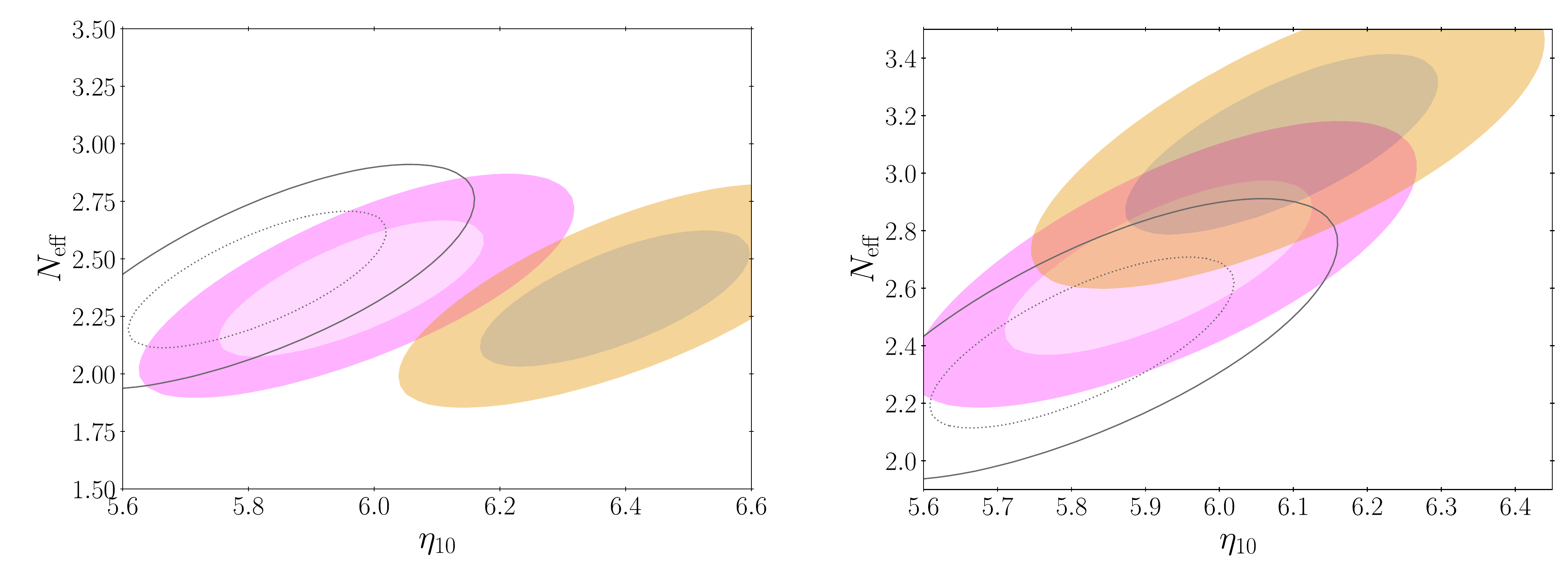}
\end{center}
\caption{\label{fig:eta_Neff_EDE} 
1$\sigma$ and 2$\sigma$ allowed regions in the $\eta_{10}$--$N_{\rm eff}$ plane from measurements of $Y_p$ and $D_p$ for the cases of EDE1 (left) and EDE2 (right). In the left panel, we take $\rho_0 = 10^{-6}~{\rm MeV}^4$ with $f_{\rm EDE} = 0.09$ (light and dark magenta for 1$\sigma$ and 2$\sigma$ regions)  and $0.50$ (light and dark orange for 1$\sigma$ and 2$\sigma$ regions). In the right panel, we take $\rho_0 = 10^{-2}~{\rm MeV}^4$ with  $f_{\rm EDE} = 0.23$ (light and dark magenta for 1$\sigma$ and 2$\sigma$ regions)  and $0.44$ (light and dark orange for 1$\sigma$ and 2$\sigma$ regions).  For reference, we also show the constraint for the case without EDE with black dotted (1$\sigma$) and solid (2$\sigma$) lines. In both panels, we assume no lepton asymmetry (i.e., $\xi_e=0$). }
\end{figure}

Next we show constraints from $Y_p$ and $D_p$ in the $\eta_{10}$--$N_{\rm eff}$ plane in the presence of EDE with $f_{\rm EDE}$ and $\rho_0$ being fixed to some values in Fig.~\ref{fig:eta_Neff_EDE}.   EDE1 and EDE2 cases are respectively depicted in left and right panels with $\xi_e=0$ fixed.  For the EDE1 case, we take $f_{\rm EDE} =0.09$ (magenta) and $0.5$ (orange) with $\rho_0 = 10^{-6}~{\rm MeV}^4$,  where light and dark regions respectively correspond to 1$\sigma$ and 2$\sigma$ allowed ones.  For the EDE2 case, $f_{\rm EDE} = 0.23$ (magenta) and $0.44$ (orange) are assumed with $\rho_0 = 10^{-2}~{\rm MeV}^4$.  

In the EDE1 case, the deuterium abundance is more affected than that of $Y_p$, and hence the effect of $f_{\rm EDE}$ degenerates with that of the baryon density since $D_p$ is sensitive to the change of $\eta_{10}$. Therefore, as $f_{\rm EDE}$ increases, the allowed region shifts almost horizontally to the right. As discussed in the previous section, when the baryon density is suggested to take a larger value, $N_{\rm eff}$ needs to be larger than the standard case and $\xi_e$ should be (positively) non-zero when no EDE is assumed (see Fig.~\ref{fig:eta_xie_Yp_Dp}). However, when EDE is present,  as the fraction of EDE increases, the allowed region is shifted to a higher value of $\eta_{10}$ although $N_{\rm eff}$ needs to be smaller than the standard value of $N_{\rm eff}= 3.046$.  Even if the $H_0$ tension really demands that we need a high value of $\eta_{10}$, the EDE1 model can reduce the discrepancy of the baryon density between the EMPRESS $Y_p$ result and CMB with a low value of $N_{\rm eff}$.

In the case of EDE2, the value of $Y_p$ is decreased by taking a larger value of $f_{\rm EDE}$, which can cancel the increase of $Y_p$ resulting from a larger value of $N_{\rm eff}$.  Therefore, by appropriately choosing the value of $f_{\rm EDE}$ and $\rho_0$, the EDE model can well fit the EMPRESS $Y_p$ \cite{Matsumoto:2022tlr} and $D_p$ result from \cite{Cooke:2017cwo} simultaneously with the standard value of $N_{\rm eff}$ and without assuming lepton asymmetry when the baryon density is $\eta_{10} \sim 6.14$ which is obtained from the Planck data in the framework of $\Lambda$CDM.  However, when a higher baryon density is suggested from CMB, which could be motivated in the light of the $H_0$ tension, one needs a larger value of $N_{\rm eff}$ than the standard one. When $\eta_{10} \sim 6.4$, the EDE2 model with $f_{\rm EDE} = 0.44$ and $\rho_0 = 8\times10^{-2}~{\rm MeV}^4$ can be well fitted to the data, but with $N_{\rm eff}=4.0$. In any case, the existence of EDE can help to improve the fit to the EMPRESS $Y_p$ result without resorting to lepton asymmetry even if a higher baryon density is suggested.

\begin{figure}[t]
\begin{center}
\includegraphics[width=16cm]{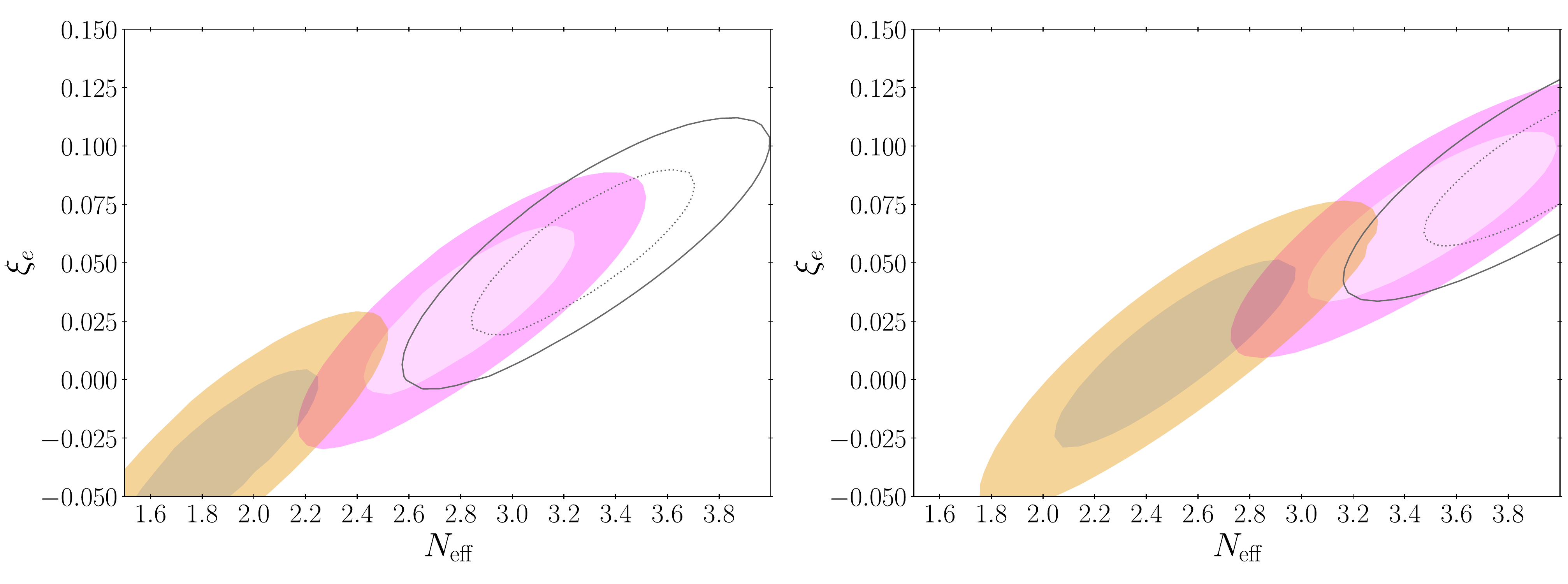}
\end{center}
\caption{\label{fig:Neff_xi_EDE1}  
1$\sigma$ and 2$\sigma$ allowed regions in the $N_{\rm eff}$--$\xi_e$ plane for the case with EDE1 adopting the priors of $\eta_{10} = \eta_{10}^{\rm ref,1}\pm \sigma_{\eta_{10}, 1}$  (left) and $\eta_{10} = \eta_{10}^{\rm ref,2}\pm \sigma_{\eta_{10}, 2}$  (right).  We take  $f_{\rm EDE} = 0.09$ (light and dark magenta for 1$\sigma$ and 2$\sigma$ regions)  and $0.44$ (light and dark orange for 1$\sigma$ and 2$\sigma$ regions)  with $\rho_0 = 10^{-6}~{\rm MeV}^4$. For reference, we also show the constraint for the case without EDE with black dotted (1$\sigma$) and solid (2$\sigma$) lines.}
\end{figure}
\begin{figure}[ht]
\begin{center}
\includegraphics[width=16cm]{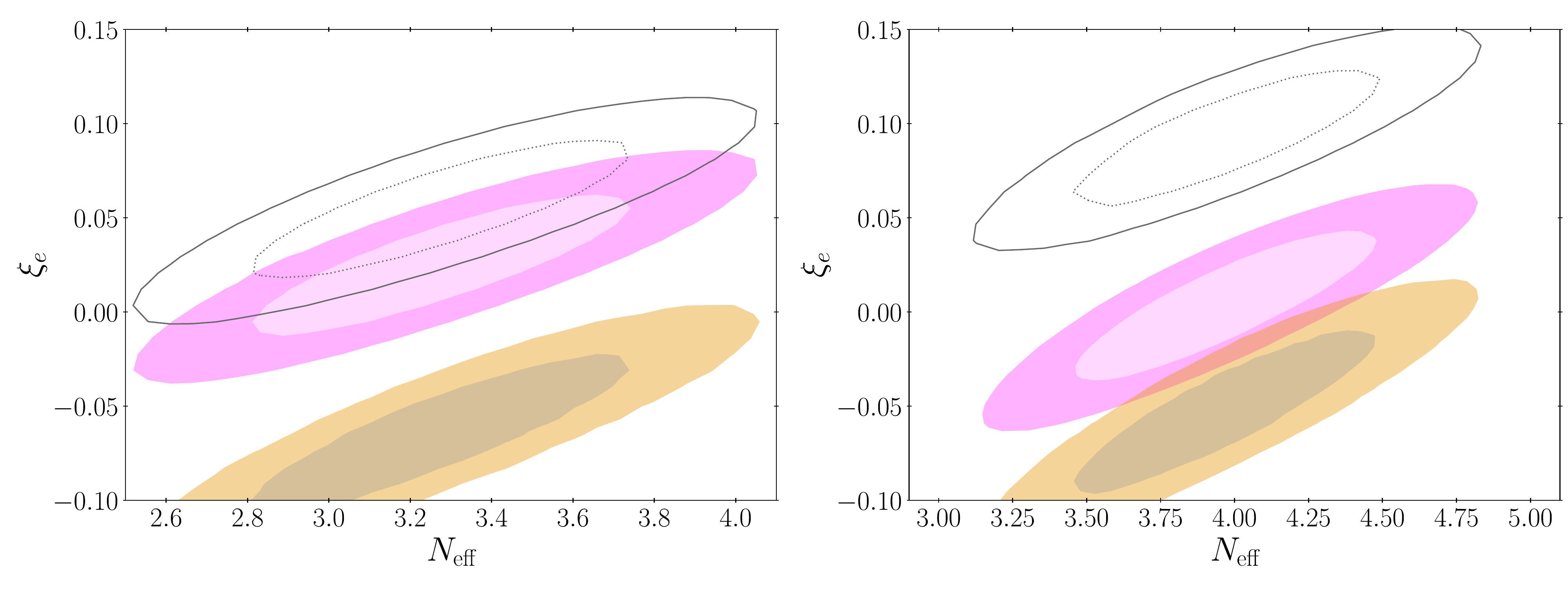}
\end{center}
\caption{\label{fig:Neff_xi_EDE2}  
1$\sigma$ and 2$\sigma$ allowed regions in the $N_{\rm eff}$--$\xi_e$ plane for the case with the EDE2,  adopting the priors of $\eta_{10} = \eta_{10}^{\rm ref,1}\pm \sigma_{\eta_{10}, 1}$  (left) and $\eta_{10} = \eta_{10}^{\rm ref,2}\pm \sigma_{\eta_{10}, 2}$  (right).  In the left panel, we take  $f_{\rm EDE} = 0.17$ (light and dark magenta for 1$\sigma$ and 2$\sigma$ regions) and $0.44$ (light and dark orange for 1$\sigma$ and 2$\sigma$ regions).  In the right panel, we take  $f_{\rm EDE} = 0.41$ (light and dark magenta for 1$\sigma$ and 2$\sigma$ regions) and $0.47$ (light and dark orange for 1$\sigma$ and 2$\sigma$ regions). 
In both panels, $\rho_0$ is assumed as  $\rho_0 = 10^{-1}~{\rm MeV}^4$. 
For reference, we also show the constraint for the case without EDE with black dotted (1$\sigma$) and solid (2$\sigma$) lines.
}
\end{figure}

Finally, we discuss constraints in the $N_{\rm eff}$--$\xi_e$ plane. 1$\sigma$ and 2$\sigma$ allowed regions are shown for the cases of EDE1 and EDE2 in Figs.~\ref{fig:Neff_xi_EDE1} and \ref{fig:Neff_xi_EDE2}, respectively.   In each figure, two priors for $\eta_{10}$, i.e., $\eta_{10}  = \eta_{10}^{\rm ref,1} \pm \sigma_{\eta_{10}, 1}$ and $\eta_{10}^{\rm ref,2}  \pm \sigma_{\eta_{10}, 2}$,  are adopted, which are respectively shown in left and right panels in the figures. For reference, we also show the constraints for the case without EDE. 

In the case of the EDE1 shown in Fig.~\ref{fig:Neff_xi_EDE1}, we take $f_{\rm EDE} =  0.09$ (magenta)  and $0.44$ (orange)  with $\rho_0 = 10^{-6}~{\rm MeV}^4$, where light and dark regions  correspond to 1$\sigma$ and 2$\sigma$ allowed ones, respectively. Since  larger $f_{\rm EDE}$ gives larger $Y_p$ and $D_p$, the decrease of $N_{\rm eff}$ and $\xi_e$ are canceled by a large value of $f_{\rm EDE}$ and hence the allowed region shifts to a lower left direction by raising $f_{\rm EDE}$. Although the standard point with $N_{\rm eff}= 3.046$ and $\xi_e=0$ is a bit away from 2$\sigma$ bound for the prior of  both $\eta_{10}^{\rm ref,1}$ and $\eta_{10}^{\rm ref,2}$, either $N_{\rm eff} = 3.046$ or $\xi_e = 0$ can be realized by appropriately choosing  the value of $f_{\rm EDE}$ and $\rho_0$ in the EDE1 case. Notice that this holds true even if the prior of a large baryon density $\eta_{10}^{\rm ref,2}$ is adopted where a more deviation from the standard values for both $N_{\rm eff}$ and $\xi_e$ are required to fit the EMPRESS $Y_p$ result without EDE. 

In the case of the EDE2 which is depicted in Fig.~\ref{fig:Neff_xi_EDE2}, we assume $f_{\rm EDE} = 0.17$ (magenta) and 0.44 (orange) in left panel. In the right panel, $f_{\rm EDE} = 0.41$ (magenta) and $0.47$ (orange) are assumed. In both panels, we take $\rho_0 = 10^{-1}~{\rm MeV}^4$.  As can be noticed from the figure, by changing the values of $f_{\rm EDE}$ and $\rho_0$, the allowed region in the $N_{\rm eff}$--$\xi_e$ plane moves almost vertically downwards. One can see that, when the baryon density is $\eta_{10} \sim 6.14$, the standard value of $N_{\rm eff}$ with no lepton asymmetry can be well fitted to the EMPRESS $Y_p$ result in the presence of EDE with appropriately chosen $f_{\rm EDE}$ and $\rho_0$. On the other hand, when a high baryon density prior of $\eta_{10}^{\rm ref,2}$ is adopted, the EMPRESS result demands the value of $N_{\rm eff}$ higher than the standard one even assuming the presence of EDE. However, it should be emphasized that both a high value of $N_{\rm eff}$ and positively non-zero $\xi_e$ are required to fit the EMPRESS $Y_p$ result when EDE is absent, which can be seen from the right panel of Fig.~\ref{fig:Neff_xie_eta}, on the other hand, once we assume EDE, one can fit the EMPRESS $Y_p$ without any lepton asymmetry, which indicates that EDE can mitigate the helium anomaly caused by the EMPRESS $Y_p$ results.

\section{Conclusion and Discussion \label{sec:conclusion}}

In this paper, first we have investigated the impact of the $H_0$ tension, in which a higher value of baryon density could be preferred than  in the $\Lambda$CDM framework, on BBN in the light of recent $Y_p$ measurement by EMPRESS.  As already pointed out \cite{Matsumoto:2022tlr}, the EMPRESS $Y_p$ result would infer a value of $N_{\rm eff}$ higher than the standard case and a positively non-zero value of $\xi_e$. However, given that models proposed to resolve the $H_0$ tension tend to prefer a higher baryon density than that in the $\Lambda$CDM adopted in  \cite{Matsumoto:2022tlr}, the deviation from the standard assumption ($N_{\rm eff} = 3.046$ and $\xi_e=0$) would be more significant.  As shown in Figs.~\ref{fig:eta_xie_Yp_Dp} and \ref{fig:Neff_xie_eta}, when a large $\eta_{10}$ is assumed, which is  suggested by the $H_0$ tension,  $N_{\rm eff}$ needs to be much larger than the standard value of $N_{\rm eff} =3.046$ and $\xi_e$ should take a large positively non-zero value to explain the EMPRESS $Y_p$ value in combination with the $D_p$ measurement of \cite{Cooke:2017cwo}.  Therefore, the $H_0$ tension could also affect BBN, which shows that the tension would have a further impact on other aspects of cosmology.

We have also studied the effects of early dark energy, which has been extensively investigated in the context of the $H_0$ tension, on the abundances of helium and deuterium. As mentioned above, the EMPRESS $Y_p$ results suggest that non-standard values of $N_{\rm eff}$ and non-zero $\xi_e$ are necessary to have a good fit to $Y_p$ and $D_p$ data simultaneously, however, by assuming the existence of early dark energy whose energy density can have a sizable fraction during BBN, the values of $N_{\rm eff}$ and/or $\xi_e$ can still take the standard values, depending on the EDE model and  parameters.  Even if a larger value of $\eta_{10}$ is assumed, which could be demanded by models resolving the $H_0$ tension,  the existence of early dark energy can still reduce the tension such that $N_{\rm eff}$ and/or  $\xi_e$ can take the values close to the standard ones. Therefore early dark energy can improve the fit even in the case of a large baryon density in the light of the EMPRESS $Y_p$ results. 

The recent EMPRESS results may suggest  physics beyond the standard cosmological model. By taking account of the $H_0$ tension, which is one of the most significant tensions in cosmology today, the modification/extension of the standard model could be more in demand. Our study  may indicate that the tensions in cosmology should  be simultaneously investigated in order to pursue a new  cosmological model, from which one can have a profound insight not only into the evolution of the Universe but also on into the fundamental physics.

\section*{Acknowledgements}
The work of T.T. was supported by JSPS KAKENHI Grant Number 19K03874.

\clearpage 
\bibliography{BBN_EDE}

\end{document}